\documentclass[journal]{IEEEtran}

\pdfoutput=1

\usepackage{cite}
\usepackage{url}

\ifCLASSINFOpdf
    \usepackage[pdftex]{graphicx}
\else
    \usepackage[dvips]{graphicx}
\fi

\usepackage{amsmath}

\begin{document}

\title{Millimeter-Wave Reflectionless Filters Using Advanced Thin-Film Fabrication}
%
%
%

\author{Matthew~A.~Morgan,~\IEEEmembership{Senior Member,~IEEE,}
        Seng Loo,
        Tod~A.~Boyd,
        and Miho Hunter
\thanks{Manuscript received 6/20/2023}
\thanks{M. Morgan and T. Boyd are with the Central Development Laboratory, National Radio Astronomy Observatory, Charlottesville,
VA, 22903 USA (e-mail: matt.morgan@nrao.edu). The National Radio Astronomy Observatory is a facility of the National Science Foundation operated under cooperative agreement by Associated Universities, Inc.}
\thanks{Seng Loo and Miho Hunter are with Anritsu Company, Morgan Hill, CA. 95037 USA.}}

\markboth{IEEE Transactions on Microwave Theory and Techniques,~Vol.~xx, No.~x, xxx~2023}%
{Morgan \MakeLowercase{\textit{et al.}}: Millimeter-Wave Reflectionless Filters Using Advanced Thin-Film Fabrication}

\IEEEpubid{0000--0000/00\$00.00~\copyright~2018 IEEE}


\maketitle

\begin{abstract}
We report on the development of millimeter-wave, lumped-element reflectionless filters using an advanced thin-film fabrication process. Based on previously demonstrated circuit topologies capable of achieving 50$\mathbf{\Omega}$ impedance match at all frequencies, these circuits have been implemented at higher frequencies than ever before by leveraging a thin-film process with better than 2 $\mathbf{\mu m}$ feature size and integrated elements such as SiN Metal-Insulator-Metal (MIM) capacitors, bridges, and TaN Thin-Film Resistors (TFRs).

\end{abstract}

\begin{IEEEkeywords}
filters, thin film circuits, reflectionless filters, absorptive filters
\end{IEEEkeywords}

\section{Introduction}

\IEEEPARstart{T}{he} concept of an absorptive filter dates back at least to the 1920's, nearly a century ago, with the work of Zobel \cite{zobel} and Bode \cite{bode}, when they were called \emph{constant-resistance networks} and were based on image-impedance principles, bridged-tee's, and cascaded, first and second-order lattice sections. However, the practical limitations on the performance and realizability of such designs precluded them from becoming widely adopted. However, interest in absorptive, or \emph{reflectionless} filters has experienced a resurgence in recent years, owing initially to the discovery of lumped-element topologies that could theoretically achieve perfect impedance match at both ports and at all frequencies from DC to infinity, with no added passband loss \cite{morgan_theoretical, morgan8392495}. At first limited to a single response type, that of a third-order Chebyshev type II filter, a broader theory of symmetric and self-dual circuit topologies was built upon this foundation until it was eventually found that any transfer function whatsoever that was realizable using lumped elements could likewise be implemented in reflectionless form \cite{morgan_artech, morgan_ladder, morgan10263592, morgan10530321, guilabert2019, khalaj-amirhosseini2016, khalaj-amirhosseini2017, lee2020}.

Seeing the advantages of a filter that absorbs stop-band energy instead of reflecting it, many researchers began searching for ways to duplicate these results using other circuit elements, such as transmission lines \cite{psychogiou2018jan, gomez-garcia2018sep, yang2020mar, gomez-garcia2020apr, wu2020aug, fan2021jan, lee2021dec}, coaxial resonators \cite{psychogiou2020aug, zhao2022}, and Surface Acoustic Wave (SAW) resonators \cite{psychogiou2019}, as well as other methodologies, most importantly coupling-routing diagrams \cite{psychogiou2020aug, zhao2022, psychogiou2019, gomez-garcia2018apr, gomez-garcia2018nov, gomez-garcia2019apr, gomez-garcia2019sep, gomez-garcia2020dec}. These often resulted in somewhat larger filters for a given frequency with limited absorption bandwidth (theoretically as well as practically), and/or with stop-band impedance-matching only at a single port. For this work, we will concentrate on lumped-element designs, as these are generally capable of the broadest absorption bandwidth in the most compact form.

Originally fabricated at relatively low frequencies using discrete surface-mount elements, these topologies were eventually implemented in the microwave regime using an Integrated Passive Device (IPD) fabrication process on GaAs wafers \cite{morgan2015}. Thus, having been reduced to practice in a form suitable for cost-effective mass manufacturing, these devices have become adopted by industry as the preferred solution in a number of commercial applications \cite{AN-75-007, AN-75-008, setty2018, shrotriya2019}. However, the difficulty of realizing good-quality lumped-elements at short wavelengths has limited their cutoff frequencies primarily to the cm-wave range (though passband and absorption bandwidths often extend much higher).

\IEEEpubidadjcol

With the advent of sophisticated thin-film fabrication that combines integrated circuit elements (MIM caps, bridges, and TFRs) with photolithography having even better resolution than is generally achievable with commercial III-V semiconductor processes, we are now prepared to implement lumped-element circuits of this kind for the first time in the millimeter-wave regime. This capability is demonstrated by the development of two prototypes: a fifth-order Chebyshev type II low-pass filter, and a seventh-order Chebyshev type II high-pass filter.

\section{Schematic Filter Design}

\begin{figure}[!t]
    \centering
    \includegraphics{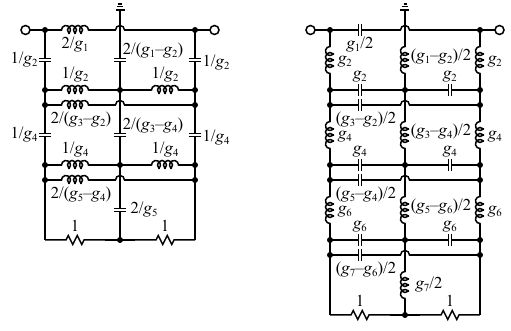}\\
    \hspace{0.1in}(a)\hspace{1.7in}(b)
    \caption{(a) Fifth-order low-pass and (b) seventh-order high-pass filters implemented in advanced thin-film technology for this paper.}
    \label{fig:schematics}
\end{figure}

Circuit diagrams of the two prototype filters reported in this paper are shown in Fig.~\ref{fig:schematics}.
Fig.~\ref{fig:schematics}(a) is a fifth-order low-pass filter, while Fig.~\ref{fig:schematics}(b) is a seventh-order, high-pass filter. Both are Chebyshev Type II designs, having equal stop-band ripple, and are theoretically reflectionless at both ports and at all frequencies.

\begin{figure}[!t]
    \centering
    \includegraphics{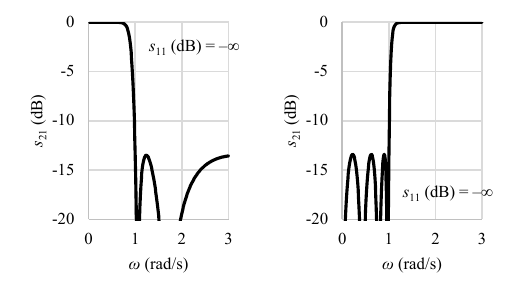}\\
    \hspace{0.3in}(a)\hspace{1.5in}(b)
    \caption{Ideal, normalized frequency response of (a) fifth-order low-pass and (b) seventh-order high-pass reflectionless filters.}
    \label{fig:ideal}
\end{figure}

The prototype parameter values used for these designs is given in Table~\ref{tab:cheb_param}.
\begin{table}[!t]
    \renewcommand{\arraystretch}{1.3}
    \caption{Chebyshev Prototype Parameters for Limiting Ripple Factors}
    \label{tab:cheb_param}
    \centering
    \begin{tabular}{@{}cc|ccccccc@{}}
        \hline\hline
        $N$ & $\varepsilon$ & $g_1$ & $g_2$ & $g_3$ & $g_4$ & $g_5$& $g_6$ & $g_7$\\
        \hline
        5 & 0.2164 & 1.337 & 1.337 & 2.164 & 1.337 & 1.337\\
        7 & 0.2187 & 1.377 & 1.377 & 2.280 & 1.498 & 2.280 & 1.377 & 1.377\\
        \hline\hline
    \end{tabular}
\end{table}
The ripple factor has been selected to achieve the maximum theoretical stop-band rejection for these topologies (without requiring transformers, which are difficult to implement at such high frequencies in planar circuit technology). That is, the ripple factor, $\varepsilon$, is given by
\begin{equation}\label{eq:ripple_factor}
    \varepsilon = \sqrt{e^{4\tanh^{-1}\left(e^{-\beta}\right)}-1} \approx \begin{cases}
        0.2164 & N=5\\
        0.2187 & N=7\\
    \end{cases}
\end{equation}
where
\begin{equation}
    \beta = 2N\sinh^{-1}\left(\sqrt{\tfrac{1}{2}\sin\left(\tfrac{\pi}{N}\right)\tan\left(\tfrac{\pi}{N}\right)}\right)
\end{equation}
and $N$ is the order of the filter \cite{morgan_artech, morgan_ladder}. Consequently, the stop-band rejection here is theoretically limited to about 13.5 dB, as shown by their ideal, normalized frequency responses in Fig.~\ref{fig:ideal}.
It has become common practice to cascade multiple stages of such designs to achieve the level of stop-band attenuation desired for a given application, however single-stage designs are sufficient for the present proof-of-concept.

\section{Physical Element Design}

Note that by selecting the minimum ripple factors, the first and last two elements in each row of Table~\ref{tab:cheb_param} are identical. This means that certain elements in the schematics of Fig.~\ref{fig:schematics} shall vanish in the final layout---e.g. the capacitor $2/\left(g_1-g_2\right)$ and the inductor $2/\left(g_5-g_4\right)$ in Fig.~\ref{fig:schematics}(a). Others will be split in two in favor of layout symmetry---for example, the series inductor having value $2/g_1$ will be implemented with two series inductors in the physical filter, mirrored, having normalized value $1/g_1$ each.

\begin{figure}[!t]
    \centering
    \includegraphics{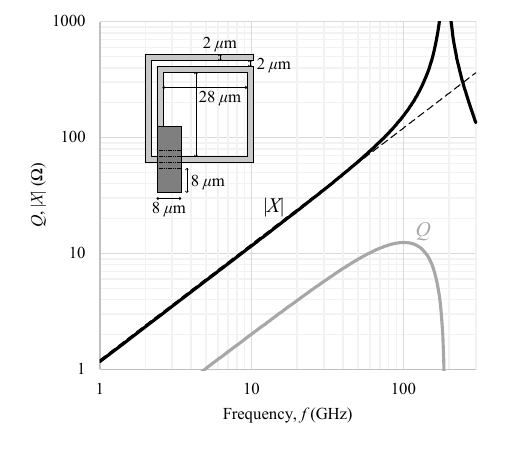}\\
    \caption{Inductor layout (inset), simulated reactance and quality factor (solid lines), and theoretical reactance of an ideal 192 pH inductor (dashed line).}
    \label{fig:inductor_sim}
\end{figure}

We selected a cutoff frequency of 60 GHz for both designs. This required capacitors ranging from 40--120 pF. These were implemented as Metal-Insulator-Metal (MIM) capacitors with a Silicon Nitride (SiN) dielectric. The inductors needed ranged from 99--199 pH. They were implemented as planar spiral coils having 1.5 turns each in the first metal layer. The fine lithography of the process made it possible to use trace widths and spacings of 2 $\mu$m each. The internal node of the coil was brought out using a dielectric bridge (the same SiN dielectric used for the capacitors). This extra parasitic capacitance at the bridge was accepted as a compromise between the simplicity of the fabrication process and the self-resonance of the inductor. A typical inductor layout for these filters is shown in Fig.~\ref{fig:inductor_sim},
as the inset on a graph showing the simulated reactance and quality factor as a function of frequency. It shows that a linear reactance curve is achieved from the lowest frequencies up to around W-band, eventually reaching a self-resonance at nearly 200 GHz. The peak $Q$, modelled at around 100 GHz, was about 12.

\section{Thin Film Fabrication Process}

The thin-film circuits were fabricated on a quartz substrate, specifically Corning Fused Silica 7980. Resistors were fabricated using Tantalum Nitride (TaN) film with a target sheet resistance of 50 $\Omega$/sq. The reflectionless filters are relatively insensitive to resistor value, so a tolerance of $\pm10$\% was accepted for this film. The first metalization comprised TiW/Au, plated up to a final thickness of 1.25~$\mu$m. Silicon Nitride (SiN) was then deposited to a target thickness of 1~$\mu$m, giving an expected capacitance density of $0.062\text{~fF}/\mu\text{m}^2$. The second metal layer again used TiW/Au, plated up to a thickness of 2.5~$\mu$m.

The final metal stack was therefore TaN/TiW/Au (1.25~$\mu$m)/SiN (1~$\mu$m)/TiW/Au (2.5~$\mu$m). An MRC 943 Series Sputter and TemesCal VES-2550 E-beam Evaporator are used for the metal deposition processes. Photolithography imaging was carried out using positive tone resists and a Canon FPA2000-il 5 X Stepper. Silicon Nitride deposition was carried out using a Novellus Concept One PECVD tool. Finally, gold electroplating utilized a Gold Sulfite plating system from Tanaka.

The potential for capacitor shorts was considered a risk for this fabrication process. The initial attempt utilized a much thinner first metalization layer of 0.4 $\mu$m, in order to keep the the surface as smooth as possible for the subsequent Silicon Nitride deposition. While the capacitor yield was excellent (no shorts were discovered), the ohmic losses of the metal traces was unacceptably high. Having greater confidence then in the capacitor process, a second run was initiated having the thicker metalization stack described above.

After all other processing steps were completed, the wafer was thinned to a final thickness of 125 $\mu$m, and diced into individual die.

\section{Measurements}

\begin{figure}[!t]
    \centering
    \includegraphics[width=2.5in]{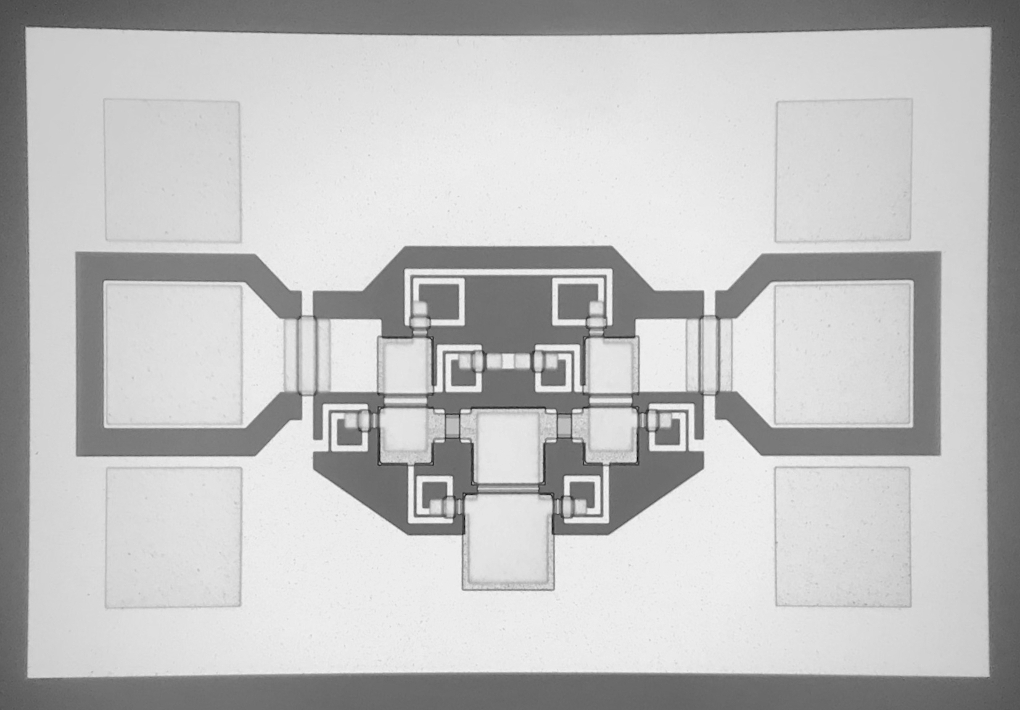}\\
    (a)\\[1em]
    \includegraphics[width=2.5in]{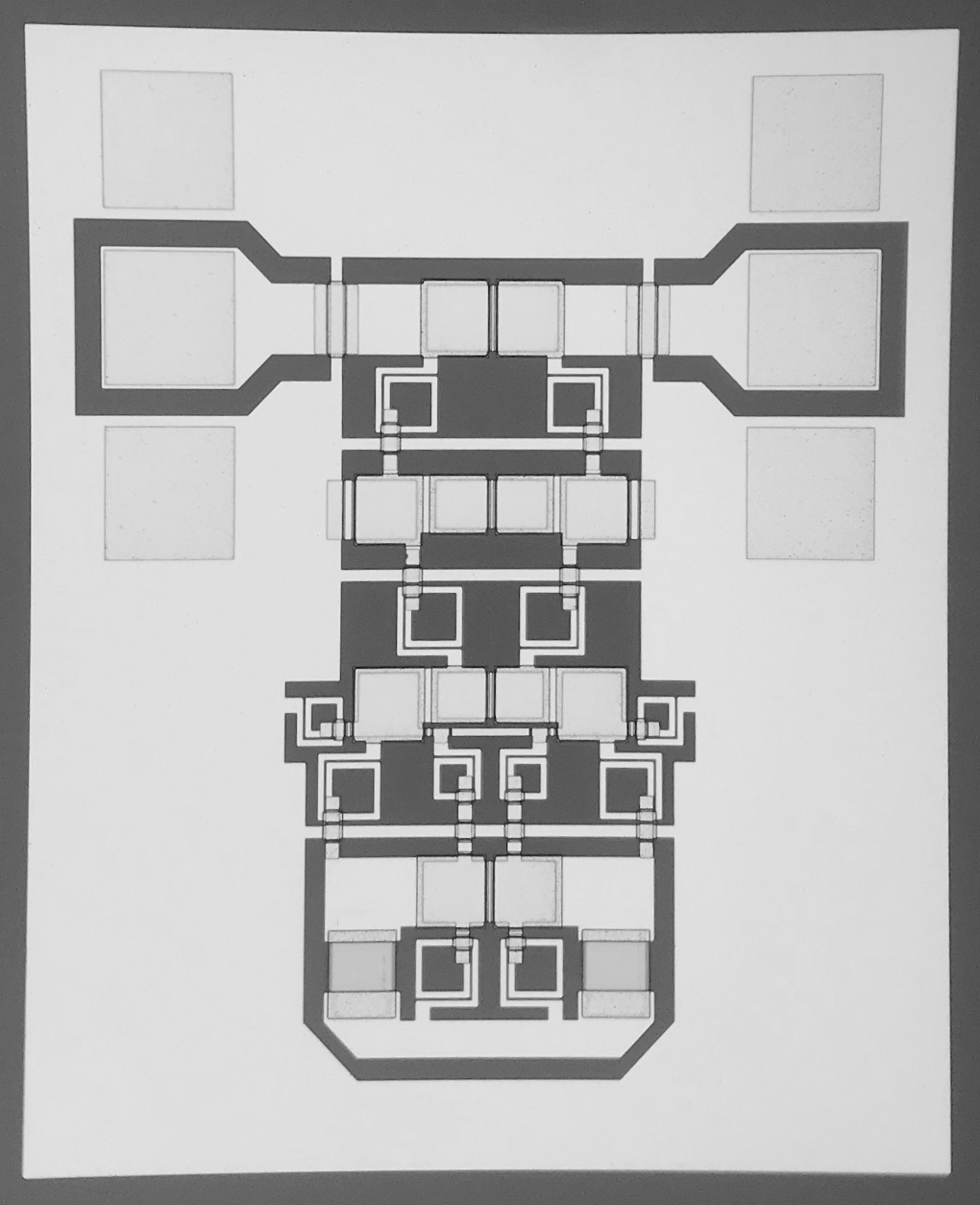}\\
    (b)\\
    \caption{Microphotographs of thin-film, reflectionless filters. (a)~Low-pass filter measuring 600~$\mu$m$\times$400~$\mu$m. (b)~High-pass filter measuring 600~$\mu$m$\times$700~$\mu$m. The substrate is 125-$\mu$m thick quartz.}
    \label{fig:photos}
\end{figure}

\begin{figure}
    \centering
    \includegraphics[width=2.5in]{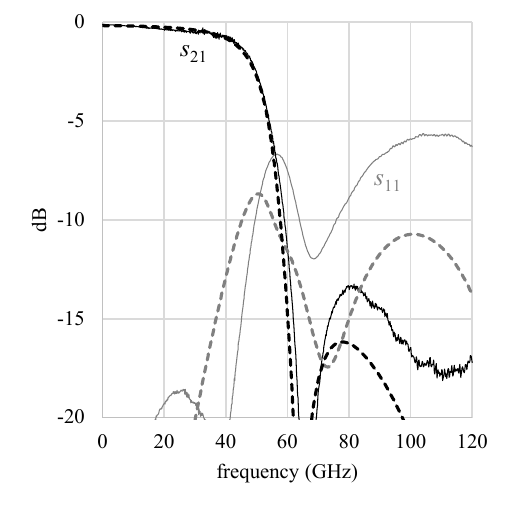}\\
    \hspace{2em}(a)\\[1em]
    \includegraphics[width=2.5in]{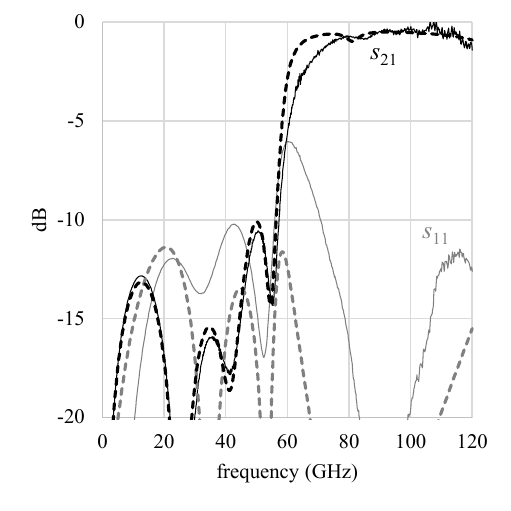}\\
    \hspace{2em}(b)\\
    \caption{Measured (solid lines) and simulated (dashed lines) s-parameters for (a)~low-pass, and (b)~high-pass reflectionless filters. The SiN dielectric was adjusted to account for an apparent shift in frequency between the measurements and the initial simulations.}
    \label{fig:measurements}
\end{figure}

Both circuits were fabricated on a 100-mm diameter quartz wafer, containing over 14,000 chips, or 7,000 of each design. Microphotographs of the two circuits are shown in Fig.~\ref{fig:photos}.
The chips were tested by wafer probe using a Keysight N5291A‐201 Vector Network Analyzer (VNA) capable of measuring two-port scattering parameters from 900~Hz--120~GHz in a single sweep. The results are plotted in Fig.~\ref{fig:measurements}.
The measured response curves for both filters, plotted in solid lines, were shifted upward in frequency 5-10\% compared to the initial simulations. This was attributed to the SiN capacitor dielectric, initially assumed to have a dielectric constant of $\varepsilon_r=7$. For the simulations plotted in Fig.~\ref{fig:measurements} (dashed lines) the dielectric constant was decreased to account for this error, but still within the acceptable range reported in literature \cite{piccirillo}. This brought the measured and simulated results within good agreement across the whole frequency range. This also partly accounted for the increased reflection coefficients in the stop-band of the low-pass filter and the transition-band of the high-pass filter.

DC resistance measurements were additionally performed at sites across the entire wafer to verify trace conductivity, and to test for shorted or leaking capacitors, which as discussed previously were considered a technical risk for this fabrication. One low-pass filter in each of 82 reticles was probe-tested for DC isolation between one of the port signal pads and ground. This is nominally expected to be an open circuit. However, if either of the first two capacitors, labeled $1/g_2$ in Fig.~\ref{fig:ideal}(a), is shorted or leaky, then a finite resistance will be measured (recall that the grounded capacitor $2/(g_1-g_2)$ vanishes since $g_1=g_2$).

\begin{figure}
    \centering
    \includegraphics[width=3.4in]{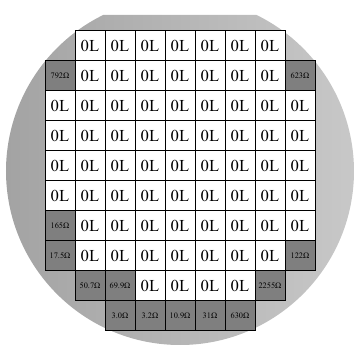}
    \caption{DC isolation between the low-pass signal port pad and ground. Out of 82 sites measured, 13 showed evidence of leakage in the capacitors.}
    \label{fig:shorts}
\end{figure}

The leakage test results are summarized in the wafer map of Fig.~\ref{fig:shorts} (``0L'' means no measurable leakage was detected).
Out of 82 reticles, 13 showed signs of some leakage, all clustered around the outer edge of the wafer. We thus estimate the global capacitor yield as $Y_C = \sqrt{69/82} \approx 92\%$ (the square root is used since two capacitors must both be good for the isolation test to pass). Closer to the center, presumably the yield is much higher (none of the devices tested in the interior, whether at DC or RF, was found to be defective).

\section{Conclusion}

This paper has presented reflectionless low-pass and high-pass filters on quartz in the millimeter-wave band. Each has a cutoff frequency at approximately 60 GHz. Furthermore, the absorptive stop-band of the low-pass filter and pass-band of the high-pass filter extend up to 120 GHz, the limit of our measurement equipment. These represent the highest operating frequencies ever reported for reflectionless filters, and a landmark achievement for strictly lumped-element-based designs.

This performance was made possible by an advanced thin-film fabrication process capable of extraordinary lithographic resolution, down to 2 $\mu$m, combined with integrated circuit elements like MIM caps, bridges, and thin-film resistors. The process shows high yield, suitable for mass manufacturing and commercialization.

\begin{figure}
    \centering
    \includegraphics{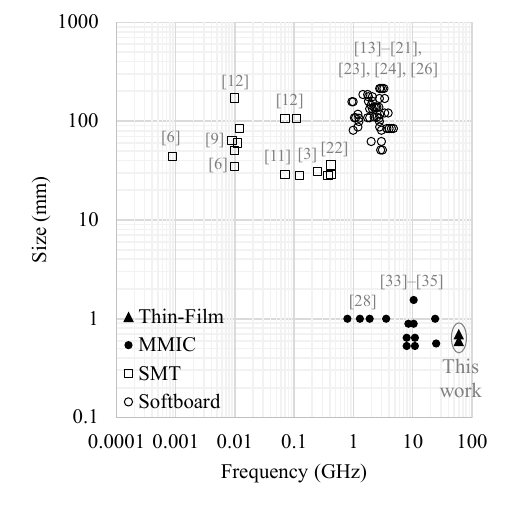}
    \caption{Comparison of this work with other approaches to reflectionless filters in terms of size (the largest linear dimension) and cutoff frequency. All cutoff corners for band-pass, band-stop, and multiple-band filters are shown. Bond pads are included in the circuit dimensions.}
    \label{fig:comparison}
\end{figure}

The advantages of these designs and of this fabrication technology are strikingly illustrated by the plot in Fig.~\ref{fig:comparison},
which compares this work to numerous others in the references in terms of compactness (on the vertical axis) versus frequency (on the horizontal). Note that both scales are logarithmic, spanning five orders of magnitude in frequency and three orders of magnitude in physical size.

It is interesting to see how well the different approaches to reflectionless filters naturally segregate themselves into clusters on such a plot. Lumped-element designs using surface-mount (SMT) components appear in the upper left as relatively large circuits at low frequency. In contrast, filters on microwave laminates or soft board materials, almost universally implemented using distributed elements such as transmission lines \cite{psychogiou2018jan, gomez-garcia2018sep, yang2020mar, gomez-garcia2020apr, wu2020aug, fan2021jan, lee2021dec} or substrate-integrated resonators \cite{psychogiou2020aug, zhao2022}, increase the frequency without any real reduction in size, thus grouping together in the upper-right corner.

The smallest and most high-frequency designs have all, until now, been implemented using MMIC fabrication, usually lumped-element, on either a GaAs IPD or Silicon CMOS wafer. The two filters reported here, using a thin-film process on quartz, appear in the extreme bottom right corner of the plot, achieving the highest frequencies in form-factors that are amongst the smallest ever reported.

\section*{Acknowledgment}
The authors with NRAO would like to thank their commercial partner, Mini-Circuits Inc., for their continued support of the development and application of reflectionless filter technology.


\bibliographystyle{IEEEtran}



\begin{IEEEbiography}[{\includegraphics[width=1in,height=1.25in,clip,keepaspectratio]{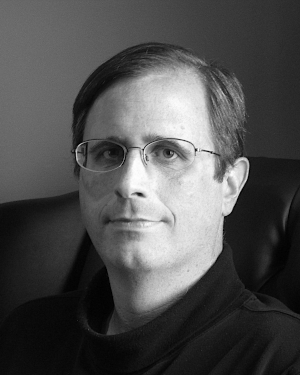}}]{Matthew A. Morgan}
(M'99--SM'17) received his B.S. in electrical engineering from the University of Virginia in 1999, and his M.S. and Ph.D. from the California Institute of Technology in 2001 and 2003, respectively.

During the summers of 1996 through 1998, he worked for Lockheed Martin Federal Systems in Manassas, VA, as an Associate Programmer, where he wrote code for acoustic signal processing, mathematical modeling, data simulation, and system performance monitoring. In 1999, he became an affiliate of NASA’s Jet Propulsion Laboratory in Pasadena, CA. There, he conducted research in the development of Monolithic Millimeter-wave Integrated Circuits (MMICs) and MMIC-based receiver components for atmospheric radiometers, laboratory instrumentation, and the deep-space communication network. In 2003, he joined the Central Development Lab (CDL) of the National Radio Astronomy Observatory (NRAO) in Charlottesville, VA, where he now holds the position of Scientist/Research Engineer. He is currently the head of the CDL’s Integrated Receiver Development program, and is involved in the design and development of low-noise receivers, components, and novel concepts for radio astronomy instrumentation in the cm-wave, mm-wave, and submm-wave frequency ranges. He has authored over 60 papers and holds twenty patents in the areas of MMIC design, millimeter-wave system integration, and high-frequency packaging techniques. He is the author of \emph{Reflectionless Filters} (Norwood, MA: Artech House, 2017).

Dr. Morgan is a member of the International Union of Radio Science (URSI), Commission J: Radio Astronomy. He received a Topic Editor's Special Mention in the IEEE THz Transactions Best Paper competition and the Harold A. Wheeler Applications Paper Award in 2015.
\end{IEEEbiography}

\begin{IEEEbiography}[{\includegraphics[width=1in,height=1.25in,clip,keepaspectratio]{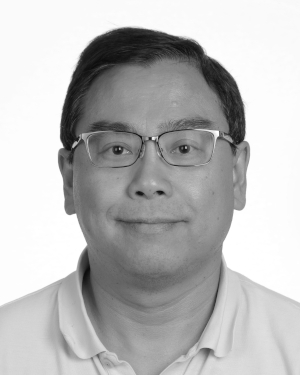}}]{Seng Loo} received his BS in Chemistry from the University of Windsor, in Ontario, Canada, and his Ph.D from Stanford University in 1991.

From 1991 through 1993 he worked as a Postdoc at the California Institute of Technology, where he carried out basic research in yeast genetics. Since 1994, he has been working in various industrial companies, including Seagate, Hyundai Electronics and Watkins-Johnson Company.

Since 2006, he has been the Fab Manager, and then Senior Business Unit Manager for the Anritsu Microelectronics Fabrication Center. In this capacity, he is tasked with leading and growing the manufacturing capabilities of the Anritsu fab. 
\end{IEEEbiography}

\vfill
\newpage 

\begin{IEEEbiography}[{\includegraphics[width=1in,height=1.25in,clip,keepaspectratio]{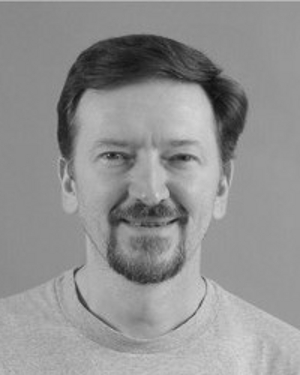}}]{Tod A. Boyd} was born in Steubenville, Ohio, in 1962. He received the A.S.E.E. degree from the Electronic Technology Institute, Cleveland, Ohio, in 1983. From 1983 to 1985, he was with Hostel Electronics in Steubenville, Ohio. In 1985, he joined Northrop Corporation’s Electronic Countermeasures Division, in Buffalo Grove, Ill., specialized in supporting the B-1B Lancer (secret clearance.) In 1990, he joined Interferometrics, Inc., Vienna, Va., where he constructed VLBA tape recorders for the international Radio Astronomy community.

Since 1996, he has been with the National Radio Astronomy Observatory’s Central Development Lab, Charlottesville, VA, where he initially assisted with the construction of cooled InP HFET amplifiers for the NASA’s Wilkinson Microwave Anisotropy Probe (WMAP) mission. Presently as a Technical Specialist IV he provides technical support for the advanced receiver R\&D initiatives. His responsibilities also include constructing low noise amplifiers for the Enhanced VLA and the Atacama Large Millimeter/sub-millimeter Array (ALMA) projects.
\end{IEEEbiography}

\begin{IEEEbiography}[{\includegraphics[width=1in,height=1.25in,clip,keepaspectratio]{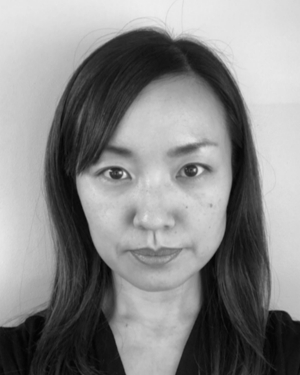}}]{Miho Hunter} joined Anritsu Company in 2009 as a Process Development Engineer, supporting the Microelectronics Fabrication Center. In 2019, she advanced to the role of Operations and Process Engineering Manager, assuming responsibility for the process engineering group tasked with designing and integrating thin film fabrication processes. Prior to joining Anritsu, Miho honed her expertise at Samsung Austin Semiconductor, specializing in the process architecture of memory devices. She holds a Master of Science degree from Stanford University.
\end{IEEEbiography}

\vfill




\end{document}